\documentclass[3p,times,twocolumn]{elsarticle}

\usepackage{ecrc}

\volume{00}

\firstpage{1}

\journalname{Nuclear Physics B Proceedings Supplement}

\runauth{}

\jid{nuphbp}

\jnltitlelogo{Nuclear Physics B Proceedings Supplement}

\usepackage{amssymb}

\biboptions{compress}

\usepackage[figuresright]{rotating}

\begin{document}

\begin{frontmatter}



\dochead{}

\title{A Hybrid Strategy for the Lattice Evaluation of the Leading
Order Hadronic Contribution to $(g-2)_\mu$}


\author{Maarten Golterman}

\address{Department of Physics and Astronomy,
San Francisco State University, San Francisco, CA 94132, USA}

\author{Kim Maltman}

\address{Department of Mathematics and Statistics, York University,
4700 Keele St., Toronto, ON CANADA M3J 1P3}
\address{and}
\address{CSSM, Univ. of Adelaide, Adelaide, SA 5005 AUSTRALIA}

\author{Santiago Peris\fnref{label1}}
\address{Department of Physics, Universitat Aut\`onoma de Barcelona,
\\ E-08193 Bellaterra, Barcelona, Spain}

\fntext[label1]{Speaker}

\begin{abstract}

The leading-order hadronic contribution to the muon anomalous magentic moment, $a_\mu^{\rm LO,HVP}$,  can be expressed as an integral over Euclidean $Q^2$ of the vacuum polarization function. We point out that a simple trapezoid-rule numerical integration of the current lattice data  is good enough to produce a result with a less-than-$1\%$ error for the contribution from the interval above $Q^2\gtrsim 0.1-0.2\ \mathrm{GeV}^2$. This leaves the interval below this value of $Q^2$ as the one to focus on in the future. In order to achieve an accurate result also in this lower window $Q^2\lesssim 0.1-0.2\ \mathrm{GeV}^2$, we indicate the usefulness of three possible tools. These are: Pad\'{e} Approximants, polynomials in a conformal variable and a NNLO Chiral Perturbation Theory representation supplemented by a $Q^4$ term.  The combination of the numerical integration in the upper $Q^2$ interval together with the use of these tools in the lower $Q^2$ interval provides a hybrid strategy which looks promising as a means of reaching the desired goal on the lattice of a sub-percent precision in the hadronic vacuum polarization contribution to the muon anomalous magnetic moment.
\end{abstract}


\end{frontmatter}


\section{Introduction}
\label{1}

Current determinations of $a_\mu=(g-2)_{\mu}/2$ in the Standard Model (SM) show a discrepancy of about $3\sigma$ with respect to experiment\cite{bnlgminus2,SMgminus2ref}. Moreover, a new Fermilab experiment expects to reduce the error by a factor of 4 in the near future which, should the central values stay the same, would mean a deviation from the SM value of about $5\sigma$! Clearly  this problem requires attention.

At present, the largest component of the error on the SM prediction is that on the leading order hadronic vacuum polarization contribution, $a_\mu^{LO,HVP}$. Since the relevant scale for this contribution is $m_\mu^2$, this requires a difficult nonperturbative calculation in QCD. Fortunately, dispersion relations allow a determination of this contribution by relating the vacuum polarization diagram to the $e^{+}e^{-}$ hadroproduction cross-section. However, the discrepancies between different experiments in the most relevant channel, $e^+e^-\rightarrow\pi^-\pi^+$, ~\cite{CMD2pipi07,SNDpipi06,BaBarpipi12,KLOEpipi12}{\footnote{A
useful overview of the experimental situation is given
in Figs.~48 and 50 of Ref.~\cite{BaBarpipi12}.}}, together with the need to maximally reduce the error in view of the coming Fermilab experiment has spurred the community to provide an independent determination of  $a_\mu^{\rm LO,HVP}$ from first principles on the lattice \cite{TB12,TB03,AB07,FJPR11,BDKZ11,DJJW12,abgp12,DPT12,FHHJPR13,FJMW13,ABGP13,BFHJPR13,gmp13,HHJWDJ13,
hpqcd14,Aubin:2012me,deRafael:2014gxa}. Such a calculation will have the extra benefit of being a very good testing ground for more difficult problems such as the light-by-light contribution to $a_\mu$, for which, at present, there are only model
estimates \cite{SMgminus2ref}.

A convenient representation for the calculation of  $a_\mu^{\rm LO,HVP}$ on the lattice is given by \cite{TB03,ER}
\begin{eqnarray}
\label{amu}
\hspace{-.7cm}a_\mu^{\rm LO,HVP}[Q^2_{min},Q^2_{max}]\!\!\!\!&=&\!\!\!\!\!-4\alpha^2\!\!\!\int_{Q^2_{min}}^{Q^2_{max}}\!\!\!\!\! dQ^2\,f(Q^2)\,
{\hat{\Pi}}(Q^2)\, , \end{eqnarray}
where $m_\mu$ is the muon mass and
\begin{eqnarray}
f(Q^2)&\!\!\!\!=&\!\!m_\mu^2 Q^2 Z^3(Q^2)\,\frac{1-Q^2 Z(Q^2)}{1+m_\mu^2 Q^2
  Z^2(Q^2)}\ ,\nonumber\\
Z(Q^2)&\!\!\!\!=&\!\!\!\left(\sqrt{(Q^2)^2+4m_\mu^2 Q^2}-Q^2\right)/
(2m_\mu^2 Q^2)\ .
\end{eqnarray}
$\hat{\Pi}(Q^2)\equiv\Pi (Q^2)-\Pi (0)$ is the subtracted polarization, defined from the hadronic electromagnetic current two-point function, $\Pi_{\mu\nu}(Q)$, via
\begin{equation}
\label{Pem}
\Pi_{\mu\nu}(Q)=\left(Q^2\delta_{\mu\nu}-Q_\mu Q_\nu\right)\Pi (Q^2)\ .
\label{polndefn}\end{equation}
The value of $a_\mu^{\rm LO,HVP}$ is obtained as $a_\mu^{\rm LO,HVP}[0,\infty]$ in Eq. (\ref{amu}). Here we restrict our
attention to the I=1 component of $\hat{\Pi}$ and denote the corresponding component of $a_\mu^{LO,HVP}$ by $\widehat{a}_\mu^{LO,HVP}$.

\begin{figure}[t]
\includegraphics[width=7cm]{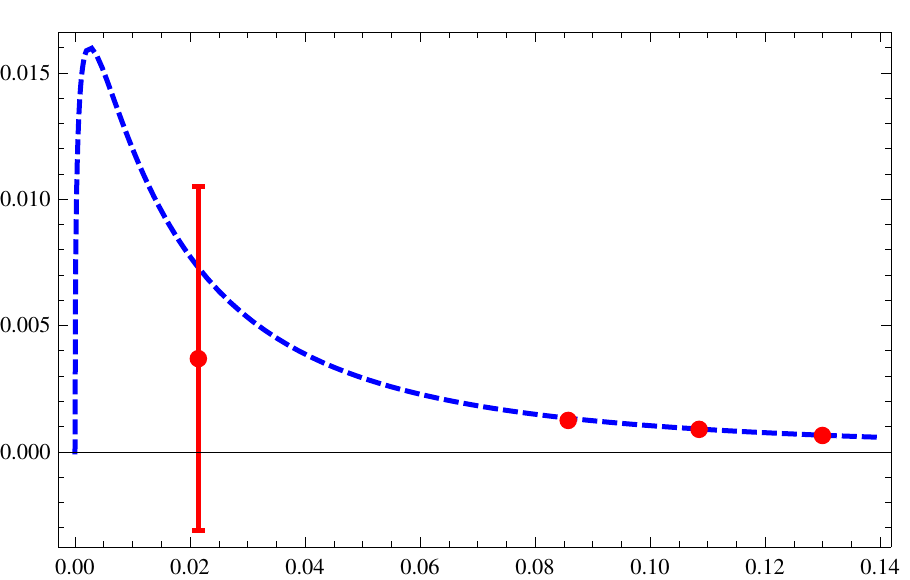}
\caption{\label{fig1} Blue-dashed curve: Integrand of Eq. (\ref{amu}) as a function of $Q^2$ in a $\tau$-data-based model. The peak is at $Q^2\sim m^2_{\mu}/4\simeq 0.003\ \mathrm{GeV}^2$.  Red data points: a set of typical lattice data.}
\end{figure}

The two-point function $\Pi_{\mu\nu}(Q)$ can in principle be computed on the lattice for non-zero $Q$  whence $\Pi (Q^2)$ can be extracted. However, this extraction is complicated in practice. A typical situation is depicted in Fig.  \ref{fig1}. In this figure, the blue dashed curve shows the result of a physically motivated model for the vacuum polarization in the $I=1$ channel based on the experimental data obtained in  non-strange hadronic $\tau$ decays,  supplemented by a successful model of duality violations at energies above the $\tau$ mass\cite{alephud,opalud99,dv72,PT,cgp08,earlydv}  (see Ref. \cite{Golterman:2014ksa} for more details). Also plotted is fake lattice data at a set of $Q_i^2$ values corresponding to a recent lattice simulation  on a $64^{3}\times 144$ lattice, with $a=0.06 fm$, $m_{\pi}=220$\ MeV and periodic boundary conditions \cite{MILC}. The fake data is obtained by letting the model $\Pi (Q_i^2)$ fluctuate according to the covariance matrix of this simulation . So, Fig. \ref{fig1} is a realistic representative picture of the situation on the lattice.

Obviously a direct numerical integration of the data shown in Fig. \ref{fig1} is not an option for an accurate evaluation of the area under the curve needed in Eq. (\ref{amu}). It is necessary to use some functional form in a fit and make sure that this functional form will faithfully reproduce the curve in Fig. \ref{fig1}, and this is why having a model becomes very useful: by knowing the answer (in the model) we can assess the size of the systematic error made. The figure of merit to keep in mind is that we need a determination of $\widehat{a}_\mu^{\rm LO,HVP}$ with better than $1\%$ precision.

One technical point: experimental spectral data can only determine the vacuum polarization through a subtracted dispersion relation. So, we can think of our model as one in which $\Pi^{I=1}(0)=0$. However, to really mimic the situation on the lattice, where the value of $\Pi^{I=1}(0)$ is unknown, we will always consider $\Pi^{I=1}(0)$ as a free parameter to be determined by fits to the data. The extent to which this value deviates from 0 then quantifies the systematic uncertainty in the determination of $\Pi^{I=1}(0)$.

\section{Hybrid strategy: ``Divide and Conquer"}
\label{2}

\begin{figure}[t]
\hspace{-.8cm}
\includegraphics[width=4.25in]{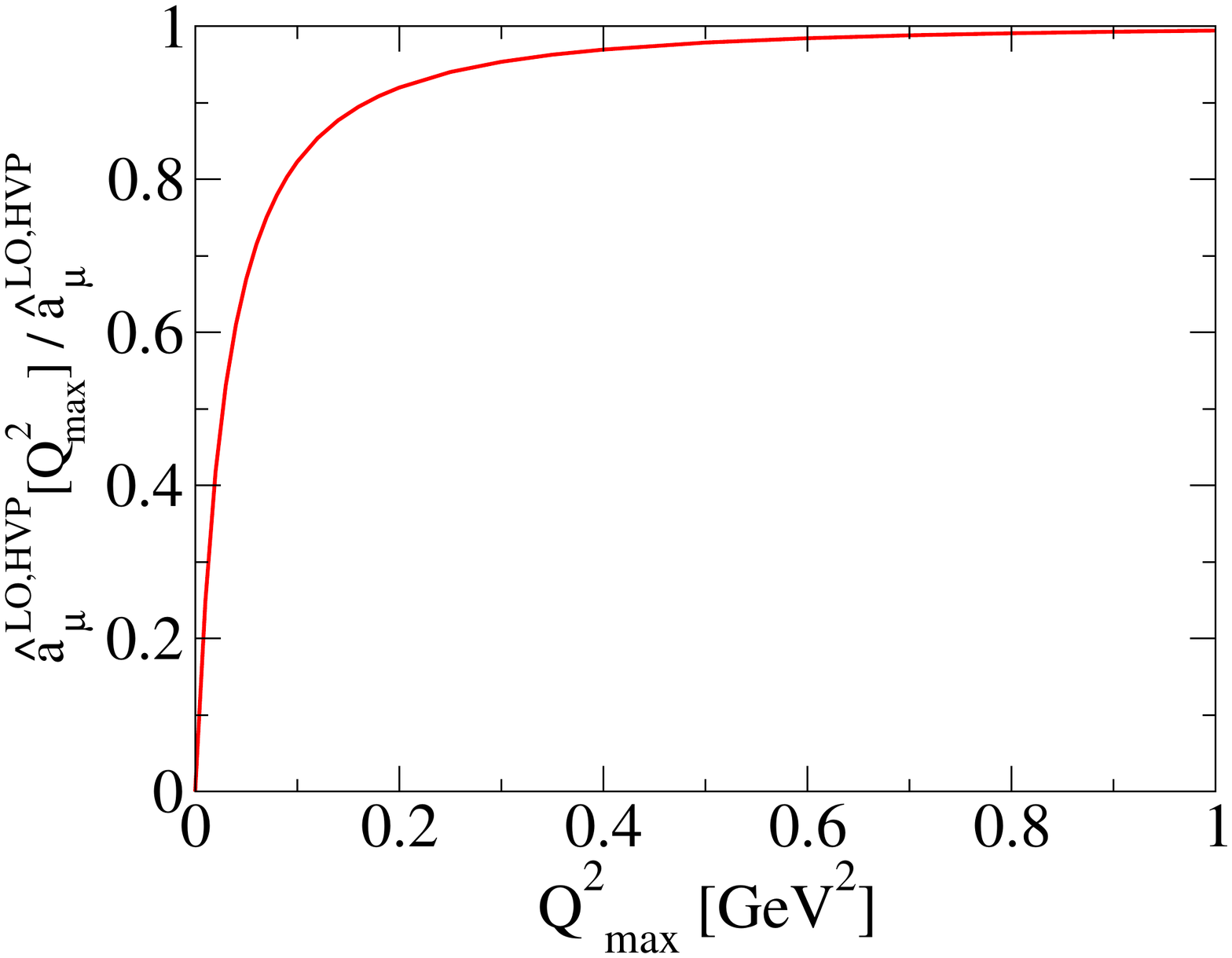}
\vspace{-1.5cm}
\caption{\label{fig2} Accumulation of the contributions to
$\widehat{a}_\mu^{\rm LO,HVP}[0,Q^2_{max}]$ as a function of $Q^2_{max}$.}
\end{figure}

\begin{figure}[t]
\hspace{-1cm}
\includegraphics[width=4.25in]
{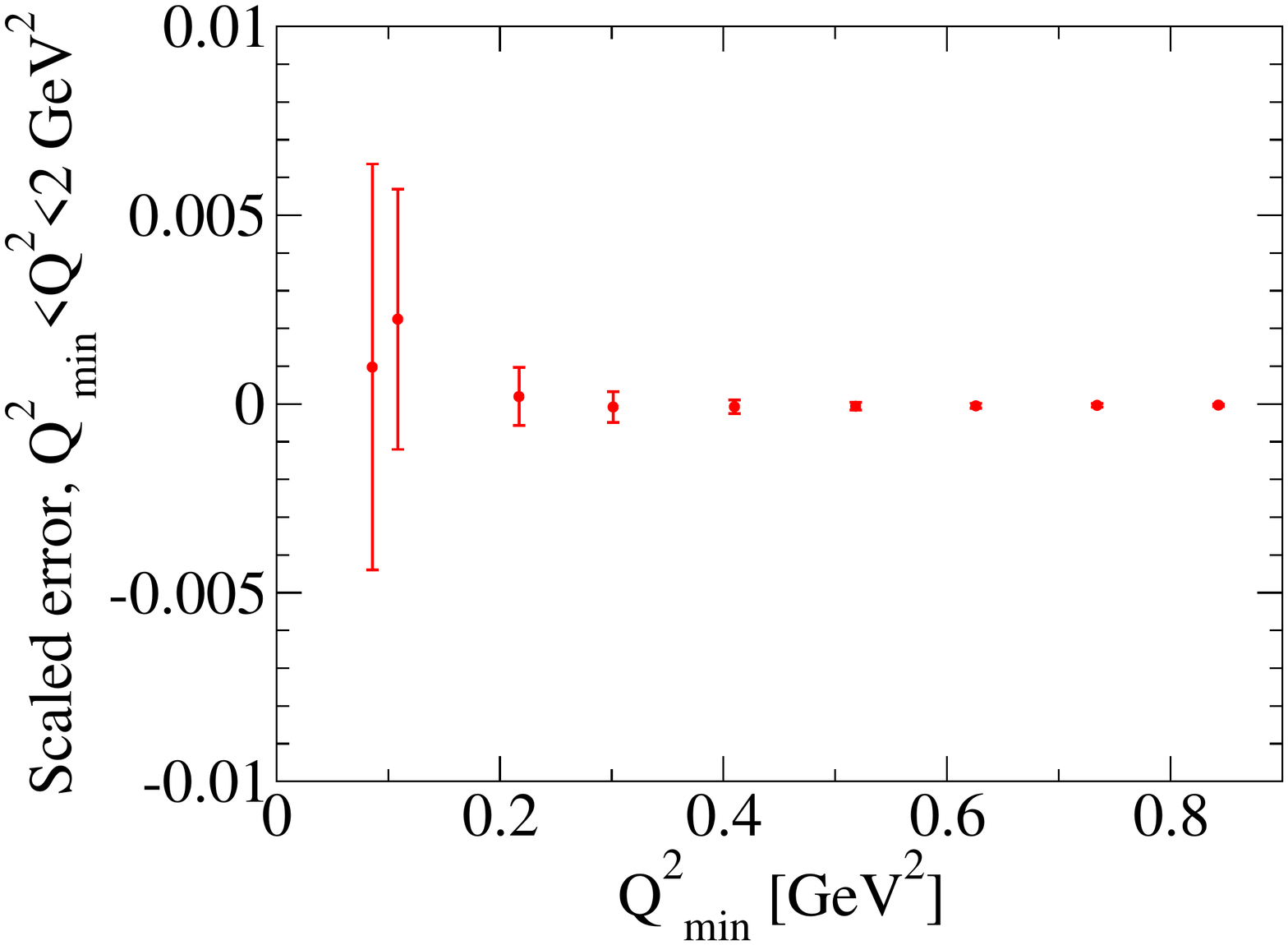}
\vspace{-1.5cm}
\caption{\label{fig3}The systematic (central values)  and statistical components (error bars)of the
error on the evaluation of $\widehat{a}_\mu^{\rm LO,HVP}[Q^2_{min},2\ \mathrm{GeV}^2]$ as a function of $Q^2_{min}$,  by direct trapezoid-rule numerical integration, as a fraction of $\widehat{a}_\mu^{\rm LO,HVP}$.}
\end{figure}

Due to the prominent peak at low $Q^2$ seen in Fig. \ref{fig1}, the integral in Eq. (\ref{amu}) is largely dominated by the contribution at low energies. This can be seen in Fig. \ref{fig2}, where more than 90\% of $\widehat{a}_\mu^{\rm LO,HVP}$ is accumulated below $Q^2\sim 0.2\ \mathrm{GeV}^2$. In fact, it is only in this low-energy region that lattice data is so scarce that  fitting is required. For higher energies a naive numerical integration of the data using  the trapezoid rule \emph{is} sufficient \cite{gmp13}. This can be seen in Fig. \ref{fig3} where, for each value of $Q^2_{min}$, we show the systematic uncertainty in $\widehat{a}_\mu^{\rm LO,HVP}[Q^2_{min},2\ \mathrm{GeV}^2]$ as a central value, and the corresponding statistical uncertainty as an error bar. Clearly, for  $Q^2_{min}\gtrsim 0.1-0.2\ \mathrm{GeV}^2$, both uncertainties are well below $1\%$. Of course, we only know the systematic uncertainty because we know the exact result in the model.

\begin{figure}[t]
\hspace{-1.cm}
\includegraphics[width=4.25in]{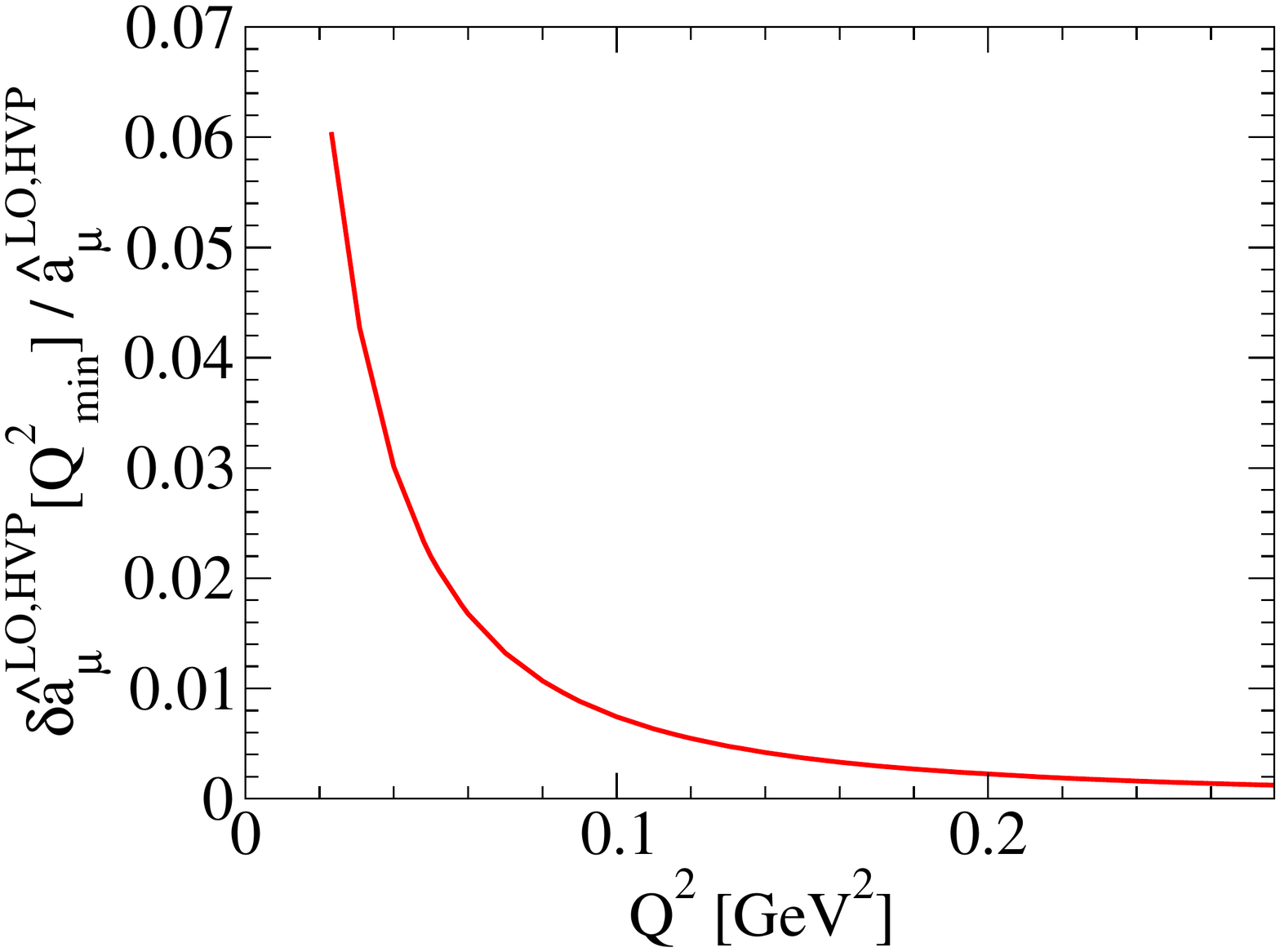}
\vspace{-1.5cm}
 \caption{\label{fig4}  The impact of an uncertainty
$\delta\Pi^{I=1}(0)=0.001$ in $\Pi^{I=1}(0)$  on
$\widehat{a}_\mu^{\rm LO,HVP}[Q^2,\infty]$
as a fraction of $\hat{a}_\mu^{\rm LO,HVP}$.}
\end{figure}

 This assumes that the subtraction constant $\Pi(0)$ is known with sufficient precision. In a previous work \cite{gmp13}, we showed that a fit of a simple $[1,1]$ Pad\'{e}\footnote{See below for further discussion on Pad\'{e}s.} to the fake data in the interval between 0 and 1 GeV$^2$ is able to determine $\Pi^{I=1}(0)$ with an uncertainty, $\delta \Pi^{I=1}(0)$, smaller than 0.001. An uncertainty $\delta \Pi^{I=1}(0)$ produces a corresponding uncertainty
\begin{equation}
\hspace{-.8cm}
\delta \widehat{a}_\mu^{\rm LO,HVP}[Q^2,\infty ]\, =\,
4\alpha^2\, \delta\Pi^{I=1}(0)\, \int_{Q^2}^\infty dQ^2\,f(Q^2)
\end{equation}
on the contribution to $\widehat{a}_\mu^{\rm LO,HVP}[Q^2,\infty]$. Fig. \ref{fig4} shows this uncertainty. We see that the error remains safely below 1\% for $Q^2 \gtrsim 0.1 -0.2\ \mathrm{GeV}^2$. This error will have to be carefully monitored in the final analysis, however, due to the rapid increase at low $Q^2$ seen in Fig. \ref{fig4}.

These observations tell us that a hybrid strategy in which one \emph{divides} the integration interval into two parts could achieve the desired precision: the first one covering $0\leq Q^2\leq 0.1-0.2\ \mathrm{GeV}^2$, and the second one covering  $0.1-0.2~\mathrm{GeV}^2\leq Q^2 < \infty$. It is only in the first part that one should fit to a functional form, while in the second part one may use a trapezoid rule integration of the data. There is no need for the long (and dangerous)  extrapolation down to $Q^2 \sim m^2_{\mu}$ from the region of ``good data'' at  $Q^2 \sim  1\ \mathrm{GeV}^2$, as has been customarily done until now.

\section{The low-$Q^2$ region: $0\leq Q^2\leq 0.1-0.2~\mathrm{GeV}^2$}
\label{3}

In this region of $Q^2$ we would like to propose a strategy based on three independent tools: Pad\'{e} Approximants, polynomials in a conformal variable and an NNLO Chiral Perturbation Theory (ChPT) representation supplemented by an analytic $Q^4$ term.

Pad\'{e} Approximants are ratios of polynomials whose coefficients are matched onto an equal number of derivatives of the original function $\widehat{\Pi}(Q^2)$ at a single point, usually at $Q^2=0$ \cite{oneptpades}, or at different values of $Q^2$ \cite{multiptpades}. In the first case they are called one-point Pad\'{e}s, while in the second case they are multi-point Pad\'{e}s. Because the vacuum polarization is a so-called Stieltjes function, there are convergence theorems which control the approximation of Pad\'{e}s to the original function everywhere in the $Q^2$ complex plane, except right on the cut of the vacuum polarization. For example, for $Q^2>0$, which is the region of interest in the integral of Eq. (\ref{amu}), one has for one-point Pad\'{e}s  that
\begin{eqnarray}
&&\hspace{-1.5cm}[1,0]_H\leq [2,1]_H\leq \cdots \leq [N+1,N]_H\leq
\hat{\Pi}^{I=1}(Q^2) \nonumber\\
&&\hspace{-1.5cm}\qquad\qquad \leq [N,N]_H\leq \cdots
\leq [2,2]_H \leq [1,1]_H\ ,\label{padeineqpr}
\end{eqnarray}
where $[M,N]_H$ represents the ratio of a polynomial of degree M over a polynomial of degree N, matched onto $M+N$ coefficients of the Taylor expansion of  $\widehat{\Pi}^{I=1}(Q^2)$ about $Q^2=0$. The subscript ``H'' refers to the fact that these Pad\'{e}s actually denote $-Q^2$ times the Pad\'{e}s conventionally used in mathematics \cite{oneptpades,multiptpades}.

In Ref. \cite{hpqcd14} it was shown how one can determine the Taylor coefficients  of $\widehat{\Pi}^{I=1}(Q^2)$ at $Q^2=0$ from time moments of the Euclidean two-point function of the vector current. To get the $n$-th term of the Taylor expansion, one needs an accurate determination of the $2 n +2$-nd time moment of the two-point function. In Ref.  \cite{hpqcd14} four Taylor coefficients were determined for $s,c$ quarks and  then Pad\'{e}s were constructed  to approximate the vacuum polarization function \emph{for all $Q^2$} and, consequently,  the integral in Eq. (\ref{amu}). The contribution for $u$ and $d$ quarks has not yet been done, and is expected to be a lot harder.

\begin{figure}[t]
\hspace{-1cm}
\includegraphics[width=4.25in]
{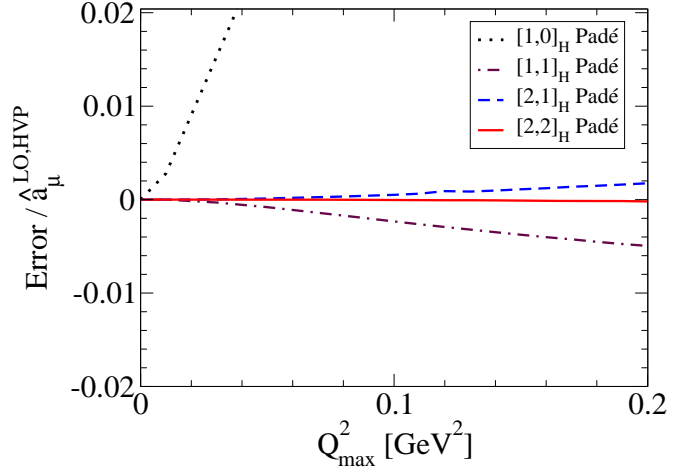}
\vspace{-1.5cm}
\caption{\label{fig5} Contribution to the error on
$\hat{a}_\mu^{LO,HVP}[0,Q^2_{max}]$ as a fraction of the full $\hat{a}_\mu^{LO,HVP}$.}
\end{figure}

However, as we have seen, it is only necessary for the Pad\'{e}s to approximate the vacuum polarization function in the low-$Q^2$ window $0\leq Q^2\leq 0.1-0.2~\mathrm{GeV}^2$. This is a big advantage as the more restricted $Q^2$ range means fewer derivatives of the correlator are required for an accurate Pad\'{e} representation in this region. Our tau-data-based model now allows us to investigate how many of these derivatives are needed to reach a given accuracy for $\widehat{a}_\mu^{\rm LO,HVP}[0,Q^2_{max}]$, as a function of $Q^2_{max}$. This is shown in Fig. \ref{fig5}. It is clear that even a $[1,1]_H$ Pad\'{e} yields an accurate enough result. In contrast, a $[2,2]_H$ Pad\'{e} would be required to reach sub-percent accuracy for the contribution on the interval $0\le Q^2\le 2$ GeV$^2$. Taking into account that the $[2,2]_H$ Pad\'{e} requires the evaluation of time moments up to the tenth order with good accuracy, whereas the $[1,1]_H$ only requires only up to the sixth order, one sees that there is a clear gain.

As an alternative to the Pad\'{e}s constructed from the Taylor expansion at $Q^2=0$, there are also the multipoint Pad\'{e}s \cite{abgp12}, whose coefficients are fixed by the value of the original function (and, if available, also its derivatives) at a discrete set of values in a $Q^2$ interval.\footnote{There are also convergence theorems for this type of Pad\'{e}s.}. This means in practice that one has to make a fit. As an example, we have done the exercise of fitting the $\tau$-based model data at the set of points $Q^2=0.10,\, 0.11,\, \cdots ,\, 0.20$~GeV$^2$ using the $[2,1]_H$
Pad\'e form, with $\hat{\Pi}^{I=1}(0)$ as a free parameter. We emphasize that this is not the fake data mentioned earlier, which was based on a real set of lattice data. With periodic boundary conditions, the current lattice data shows too large errors and a too small number of values of $Q^2$ for this type of fit to be successful.  So, even though the present exercise cannot be considered as fully realistic now, it may become feasible in the future thanks to error-reduction techniques \cite{bis12,amaref}  and new theoretical ideas \cite{DJJW12,DPT12,FHHJPR13,FJMW13,ABGP13,lehnerlattice2014}. Even so, we find that it is necessary to go to the $[2,1]_H$ Pad\'{e} to get down to the sub-percent level in the systematic error of the integral in Eq. (\ref{amu}) from 0 to 0.1-0.2 GeV$^2$. The rule of thumb is that, in order to achieve the same level of accuracy, Pad\'{e}s constructed from fitting in an interval require one order more than those obtained by the Taylor coefficients at the origin. See Ref. \cite{Golterman:2014ksa} for more details.

Another possibility is a conformal expansion of the subtracted vacuum polarization. The Taylor expansion of $\hat{\Pi}^{I=1}(Q^2)$ converges only for $\vert Q^2\vert <4m_\pi^2$, which is too small to be useful. The convergence properties can be improved by going to the conformal variable
\begin{equation}
\label{exppar}
w(Q^2)=\frac{1-\sqrt{1+z}}{1+\sqrt{1+z}}\ ,\qquad z=\frac{Q^2}{4m_\pi^2}\ ,
\end{equation}
and then expanding in $w$. The new function $\hat{\Pi}^{I=1}(Q^2(w))$ should converge faster because the whole $Q^2$ complex plane is mapped onto the unit disc, with the $Q^2$ cut at the boundary. The region of convergence now includes those $w$ corresponding to the whole positive real $Q^2$ axis, which is what is needed in Eq. (\ref{amu}).

\begin{figure}[t]
\hspace{-1cm}
\includegraphics[width=4.25in]
{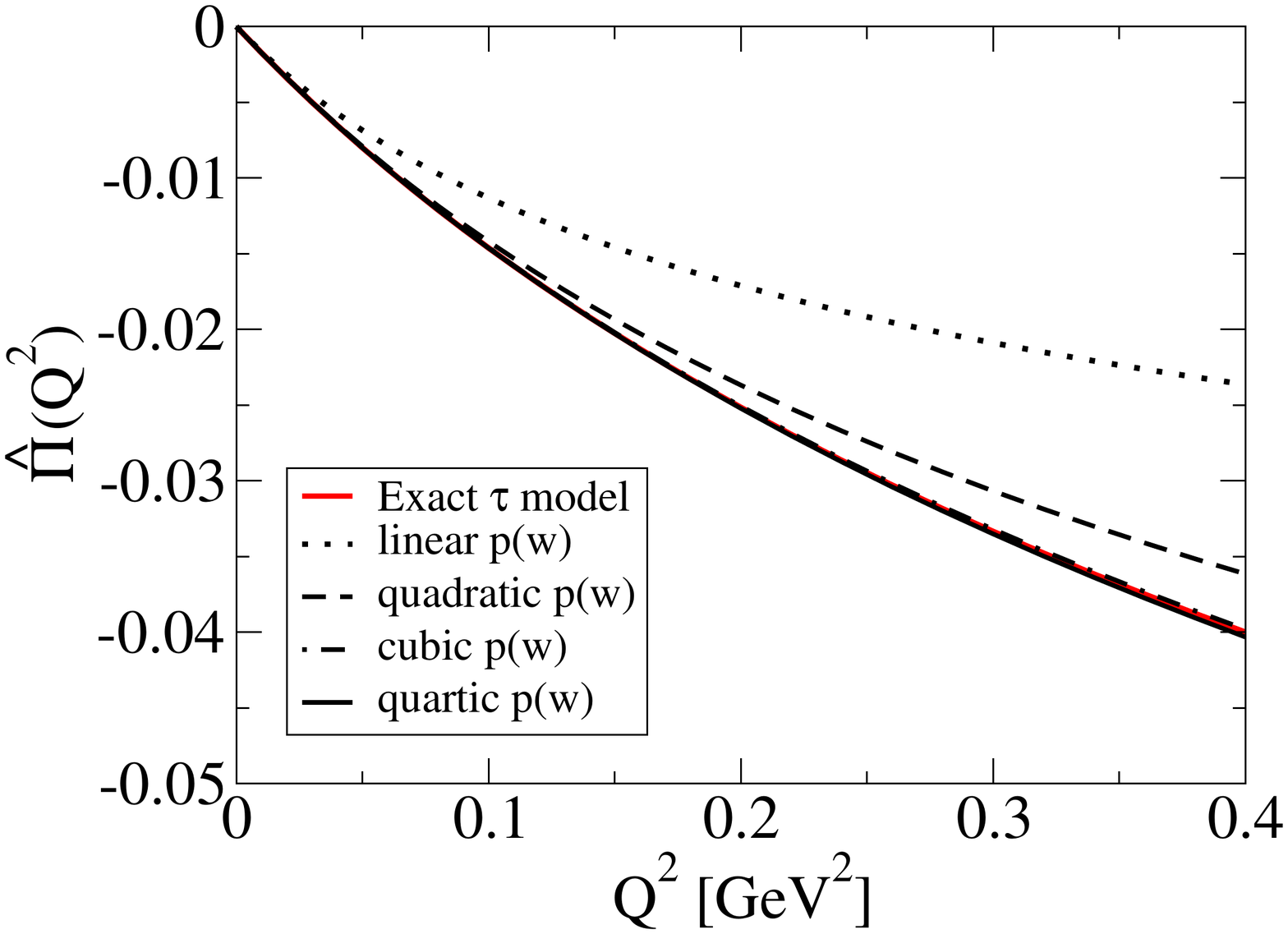}
\vspace{-1.5cm}
\caption{\label{fig6}Comparison of the results of the conformal
polynomial
representations up to quadratic order with the exact
$\tau$-data-based model  for  $\hat{\Pi}^{I=1}(Q^2)$. }
\end{figure}

As in the case of the Pad\'{e} Approximants, the conformal expansion can be constructed from the lowest Taylor coefficients of the vacuum polarization function which, in turn, can be determined by the Euclidean time moments of the two-point correlator. With up to the fourth order Taylor coefficient, one can construct a linear, quadratic, cubic and quartic polynomial approximations to $\hat{\Pi}^{I=1}(Q^2)$ and compare with the exact result. The result of this comparison is shown in Fig. \ref{fig6}.

This figure shows that the linear polynomial is clearly insufficient. The quadratic version
yields much better estimates for $\hat{a}_\mu^{\rm LO,HVP}[0,Q^2_{max}]$, i.e.
$0.6\%$ and $1\%$ below the exact model values for $Q^2_{max}=0.1$ and
$0.2$~GeV$^2$, respectively. In the case of the
cubic representation, the corresponding errors  are $0.02\%$ and $0.04\%$.  These numbers are to
be compared to $0.3\%$ and $0.5\%$ for the $[1,1]_H$ Pad\'e
(which has the same number of parameters as the quadratic polynomial),
and $0.06\%$ and $0.2\%$ for the $[2,1]_H$ Pad\'e (which has
same number of parameters as the cubic polynomial).

Also, as in the case of the Pad\'{e}s, one may consider fitting these conformal polynomials in a interval of $Q^2$ values to avoid the difficulties of the Euclidean time moment calculation. We have repeated the same exercise we carried out for the Pad\'{e}s and the $\tau$-data-based model above in the interval $Q^2=0.10,\, 0.11,\, \cdots ,\, 0.20$~GeV$^2$ and find that, again,  polynomials in the conformal variable obtained from fitting require one order more than those obtained from the Taylor coefficients at $Q^2=0$ in order to reach the same accuracy for $\hat{a}_\mu^{\rm LO,HVP}[0,0.1-0.2~\mathrm{GeV}^2]$.

 Finally, we would like to comment on a strategy based on ChPT \cite{gl84,gl85}. Since the region of interest is given by low values of $Q^2$, $0\leq Q^2\leq 0.1-0.2~\mathrm{GeV}^2$, in principle ChPT should yield a good representation. However, the highest order of ChPT available is NNLO \cite{gk95,abt00} and, unfortunately,  this turns out to be insufficient. This is not very surprising: the old success of phenomenological descriptions like Vector Meson Dominance suggest that vector resonances have to play a very important role. However, NNLO is just the first order at which vector mesons appear in the subtracted vacuum polarization function through the associated low-energy constants (LECs).  Estimating the size of $\rho$-induced corrections beyond NNLO, one finds NNNLO corrections, for which no complete calculation exists, are likely to become important already for $Q^2\sim 0.1$ GeV$^2$.

Given this state of affairs, as an exploratory exercise, we have supplemented the NNLO result with an extra  $Q^4$ term to construct an \emph{incomplete} NNNLO form which we call NN$^{\prime}$LO. The idea is that we want to know whether, with sufficiently good low-$Q^2$ data, at least this form is capable of representing  ${\hat{\Pi}^{I=1}}(Q^2)$ accurately enough to produce a sub-percent result for $\hat{a}_\mu^{\rm LO,HVP}[0,0.1-0.2 ~\mathrm{GeV}^2]$.

\begin{figure}[t]
\hspace{-.8cm}
\includegraphics[width=4.2in]
{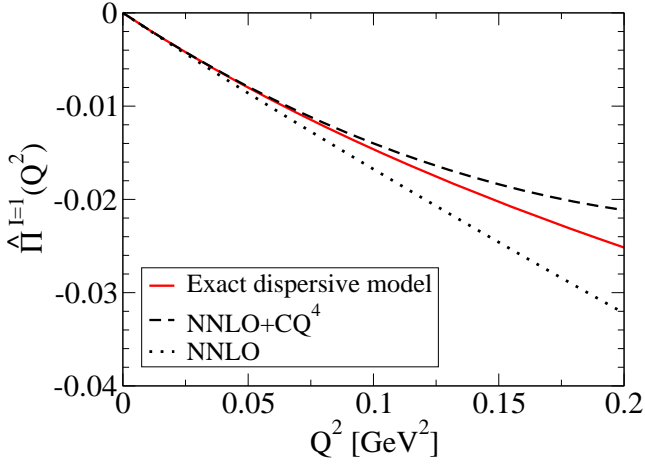}
\vspace{-1.5cm}
\caption{\label{fig7}Comparison of the results of the NN$^\prime$LO
representation
and the $\tau$-data-based model for $\hat{\Pi}^{I=1}(Q^2)$ (red solid curve).
The dashed line shows the result including the phenomenological
term $C Q^4$, where $C$ is a constant, the dotted line the result
with this term removed.}
\end{figure}

Using again our $\tau$-data based model we can determine the unknown LECs needed to construct our NN$^{\prime}$LO representation from the first and second derivatives of the vacuum polarization function in this model, and then compare to the exact result for  ${\hat{\Pi}^{I=1}}(Q^2)$ (see Ref. \cite{Golterman:2014ksa} for more details). The result is shown in Fig. \ref{fig7}. As one can see, the NNLO dotted curve fails to give an accurate representation of the red solid curve, representing the exact result, in the region of interest.

For $Q^2_{max}=0.1$~GeV$^2$, we find that our NN$^{\prime}$LO ChPT value for $\hat{a}_\mu^{\rm LO,HVP}[0,Q^2_{max}]$
 is $0.6\%$ below the exact value, while for $Q^2_{max}=0.2$~GeV$^2$, it is $1.4\%$ below.
Although the value at $Q^2_{max}=0.1$~GeV$^2$ is acceptable, this is
clearly worse than the approximation obtained using a $[1,1]_H$ Pad\'e, which was also
determined from the first and second derivatives at $Q^2=0$. For comparison, NNLO ChPT, which
corresponds to removing the extra $Q^4$ term, yields values $4\%$ and $18\%$ above
the exact value, at $Q^2_{max}=0.1$ and $0.2$~GeV$^2$, respectively. This is clearly insufficiently accurate.

As before, we can also attempt to construct our NN$^{\prime}$LO function from a fit to a set of $Q^2$ values in an interval, instead of from derivatives at $Q^2=0$. However, Fig. \ref{fig7} shows that, unlike the case of Pad\'{e}s, for which a fit in the interval $[0.1,  0.2]\ \mathrm{GeV}^2$ was in principle possible, in the case of the NN$^{\prime}$LO representation the fit window would have to be \emph{below} $Q^2 \sim 0.1\ \mathrm{GeV}^2$. Accurate lattice data for a dense set of such low values of $Q^2$ may be harder to get. We conclude that a ChPT-based approach may be useful for some sort of consistency check but it is unlikely to be as useful as Pad\'{e}s or the polynomials in the conformal variable we have described.

\section{Errors for the hybrid strategy and conclusions}
\label{4}

The contribution from the vacuum polarization function  to the muon $ \widehat{a}_\mu^{\rm LO,HVP}$, Eq. (\ref{amu}), requires knowledge of this function for all values of $Q^2$. Even though for $Q^2$ large enough, say $Q^2 \gtrsim 2 \ \mathrm{GeV}^2$,  one may apply perturbation theory, the largest contribution comes from values of $Q^2\sim m^2_{\mu}/4$ which are much smaller. Lattice data provides a discrete set of points  in this region of small $Q^2$ but, currently, neither the accuracy nor the density of these points is sufficient to allow a numerical integration to  cover the region of the integral going from $Q^2 \sim 2 \ \mathrm{GeV}^2$ down to $Q^2=0$. The standard method of calculation until now has been to use an extrapolation of  the data from $Q^2 \sim 1-2\ \mathrm{GeV}^2$, where the data is quite accurate, all the way down to $Q^2=0$. This results in an unwanted systematic error which makes a reliable sub-1\% precision in the total contribution an impossible goal to reach.

In Ref. \cite{Golterman:2014ksa} we have pointed out that it is advantageous to divide the integration region into two parts, one covering $0.1-0.2\ \mathrm{GeV}^2 \lesssim Q^2 \lesssim 2\ \mathrm{GeV}^2$ and another one covering $0 \lesssim Q^2 \lesssim 0.1-0.2\ \mathrm{GeV}^2$. The contributions from these two regions can then be evaluated using a hybrid strategy. First, for the upper part,  one can apply a simple trapezoid-rule numerical integration. Existing lattice data is already good enough to produce a value for this part of the integral which is sufficiently accurate. The problematic region is the lower part, where current lattice data shows large errors and the $Q^2$ coverage is too sparse to allow a numerical integration.

We have pointed out that there are three methods likely to have some utility in dealing with the region  $0 \lesssim Q^2 \lesssim 0.1-0.2\ \mathrm{GeV}^2$. These are Pad\'{e}s, polynomials in the conformal variable and ChPT. Of these, we have seen that Pad\'{e}s and the polynomials in the conformal variable will probably be the most efficient, while ChPT, because it is only fully known up to NNLO, will not be so optimal.

The Pades and polynomials in the conformal variable can both be constructed from the values of the derivatives of the
subtracted polarization with respect to $Q^2$ at $Q^2=0$, values which can, in principle, be determined from the Euclidean time moments of the vector current correlator, as Ref. \cite{hpqcd14} has shown for $s$ and $c$ quarks. A question that still remains is whether such a determination will be precise enough for the case of $u$ and $d$ quarks. Alternatively, one could also construct these Pad\'{e}s and polynomials in the conformal variable by fitting lattice data in a subset of points in the region $0 \lesssim Q^2 \lesssim 0.1-0.2\ \mathrm{GeV}^2$, once the data in this region becomes better in the future.

In order to understand how precise the values of ${\hat{\Pi}^{I=1}}(Q^2)$ and its derivatives at $Q^2=0$ have to be to reach the desired sub-percent total error, we can construct the Pad\'{e} $[1,1]_H$ (which was found to be sufficient
to reach this precision in the low-$Q^2$ region) as
\begin{equation}
\hat{\Pi}(Q^2)\,=\,\Pi (Q^2)-\Pi (0)\,  =\, {\frac{a_1Q^2}{1+b_1Q^2}}\, .
\label{pade11form}
\end{equation}
Errors $\delta a_1$ and $\delta b_1$
on the parameters $a_1$ and $b_1$ produce
associated errors
\begin{eqnarray}
&&\hspace{-1.5cm}\delta_{a_1}\widehat{a}_\mu^{LO,HVP}[Q_{min}^2]=\!\!
-4\alpha^2\!\!\! \int_0^{Q_{min}^2}\!\!\!\!\!\!\!\!dQ^2\, f(Q^2)\, \left(
{\frac{Q^2}{1+b_1Q^2}}\right)\delta a_1\ ,\nonumber\\
&&\hspace{-1.5cm}\delta_{b_1}\widehat{a}_\mu^{LO,HVP}[Q_{min}^2] =
4\alpha^2\!\!\! \int_0^{Q_{min}^2}\!\!\!\!\!\!\!\!dQ^2\, f(Q^2)\, \left(
{\frac{a_1Q^4}{(1+b_1Q^2)^2}}\right)\delta b_1\ ,\nonumber \\
\label{pade11errors}\end{eqnarray}
on $\widehat{a}_\mu^{\rm LO,HVP}[0,Q_{min}^2]$. Using our $\tau$-data-based model we have found that a sub-percent error on $a_1$, together with at most a few percent error on $b_1$, will be enough to obtain a sub-percent error on   $\widehat{a}_\mu^{\rm LO,HVP}[0,0.1-0.2\ \mathrm{GeV}^2]$.

Further quantitative studies using our $\tau$-based model will become possible once improved lattice data becomes available. This will allow us to construct fake data sets with realistic errors and correlations from the point of view of the lattice, which can then be used to assess the systematic error associated with the use of fit forms in the low $Q^2$ region of any lattice evaluation of the contribution from the hadronic vacuum polarization to the muon anomalous magnetic moment.

\vspace{.2cm}

MG is supported in part by the US Department of Energy, KM is supported by the Natural Sciences and Engineering Research Council of Canada,
and SP is supported by CICYTFEDER-FPA2011-25948, 2014 SGR 1450, and
the Spanish Consolider-Ingenio 2010 Program CPAN (CSD2007-00042).








\end{document}